\documentclass[11pt]{article}

\usepackage{acl}

\usepackage{times}
\usepackage{latexsym}

\usepackage[T1]{fontenc}
\usepackage[utf8]{inputenc}

\usepackage{microtype}

\usepackage{graphicx}
\graphicspath{{./figs/}}
\usepackage{booktabs}

\usepackage{xcolor}

\usepackage{xspace}

\newcommand{\LSTMmodelname}{{\it ContentOnly}\xspace} 

\newcommand{\hide}[1]{}
\newcommand{\etc}{\textit{etc.}\xspace}
\newcommand{\eg}{\textit{e.g.,}\xspace}
\newcommand{\ie}{\textit{i.e.,}\xspace}
\newcommand{\sect}{\S}
\newcommand{\reddit}{reddit\xspace}

\newcommand{\ignore}[1]{}
\newcommand{\etal}{\textit{et al.}\xspace}

\usepackage{tabularx}
\usepackage{multirow, multicol}
\newcommand*\rot{\rotatebox{90}}

\usepackage{colortbl}

\usepackage{pgfplots}  
\usepackage{tikz}
\pgfplotsset{compat=1.16}
\definecolor{tab_blue}{RGB}{31,119,180}
\definecolor{tab_orange}{RGB}{255,127,14}
\definecolor{tab_green}{RGB}{44,160,44}
\definecolor{tab_red}{RGB}{214,39,40}
\definecolor{tab_purple}{RGB}{148,103,189}
\definecolor{tab_brown}{RGB}{140,86,75}
\definecolor{tab_pink}{RGB}{227,119,194}
\definecolor{tab_gray}{RGB}{127,127,127}
\definecolor{tab_olive}{RGB}{188,189,34}
\definecolor{tab_cyan}{RGB}{23,190,207}

\definecolor{firebrick}{RGB}{178,34,34}
\definecolor{silver}{RGB}{192,192,192}
\definecolor{dodgerblue}{RGB}{30,144,255}
\definecolor{darkblue}{RGB}{0,0,139}
\definecolor{darkmagenta}{RGB}{139,0,139}
\definecolor{fuchsia}{RGB}{255,0,255}
\definecolor{teal}{RGB}{0,128,128}
\definecolor{springgreen}{RGB}{0,255,127}
\definecolor{limegreen}{RGB}{50,205,50}
\definecolor{darkgreen}{RGB}{0,100,0}
\definecolor{peru}{RGB}{205,133,63}
\definecolor{saddlebrown}{RGB}{139,69,19}
\definecolor{black}{RGB}{0,0,0}

\definecolor{mean_frac_extreme_right_bias}{rgb}{1.0, 0.0, 0.0}
\definecolor{mean_frac_center_bias}{rgb}{0.5019607843137255, 0.0, 0.5019607843137255}
\definecolor{mean_frac_center_left_bias}{rgb}{0.5294117647058824, 0.807843137254902, 0.9215686274509803}
\definecolor{mean_frac_left_bias}{rgb}{0.0, 0.0, 1.0}
\definecolor{mean_frac_very_low_factual}{rgb}{1.0, 0.27058823529411763, 0.0}
\definecolor{mean_frac_low_factual}{rgb}{1.0, 0.6470588235294118, 0.0}
\definecolor{mean_frac_mixed_factual}{rgb}{1.0, 0.8431372549019608, 0.0}
\definecolor{mean_frac_mostly_factual}{rgb}{0.6784313725490196, 1.0, 0.1843137254901961}
\definecolor{mean_frac_high_factual}{rgb}{0.19607843137254902, 0.803921568627451, 0.19607843137254902}
\definecolor{mean_frac_very_high_factual}{rgb}{0.0, 0.5019607843137255, 0.0}
\definecolor{mean_num_pnnl_deceptive}{rgb}{1.0, 0.7529411764705882, 0.796078431372549}
\definecolor{mean_frac_pnnl_deceptive}{rgb}{0.4, 0.2, 0.6}
\definecolor{frac_true_deceptive}{rgb}{1.0, 0.0, 1.0}

\renewcommand{\baselinestretch}{.95}

\usepackage{enumitem}
\setlist{nosep} 

\title{Leveraging Community and Author Context to  
Explain the Performance and Bias of Text-Based Deception Detection Models}

\author{
Galen Weld\textsuperscript{1}, Ellyn Ayton\textsuperscript{2}, Tim Althoff\textsuperscript{1}, Maria Glenski\textsuperscript{2}\\
\normalsize{\textsuperscript{1}{\normalfont Paul G. Allen School of Computer Science and Engineering, University of Washington}} \\
\normalsize{\textsuperscript{2}{\normalfont National Security Directorate, Pacific Northwest National Laboratory}} \\
\normalsize{\normalfont \{gweld, althoff\}@cs.washington.edu, first.last@pnnl.gov }
}

\begin{document}
\maketitle

\begin{abstract}

Deceptive news posts shared in online communities can be detected with NLP models, and much recent research has focused on the development of such models.  
In this work, we use characteristics of online communities and authors --- the context of how and where content is posted --- to explain the performance of a neural network deception detection model and identify sub-populations who are disproportionately affected by model accuracy or failure. We examine \textbf{\it who} is posting the content, and \textbf{\it where} the content is posted to.
We find that while author characteristics are better predictors of deceptive content than community characteristics, both characteristics are strongly correlated with model performance. 
{Traditional performance metrics such as F1 score may fail to capture poor model performance on isolated sub-populations such as specific authors, and as such, more nuanced evaluation of deception detection models is critical.}

\end{abstract}

\section{Introduction}\label{sec:intro}

The spread of deceptive news content in online communities significantly erodes public trust in the media~\cite{barthel_mitchell_holcomb_2016}. Most social media users use these platforms as a means to consume news -- 71\% of Twitter users and 62\% of Reddit users -- and in general, 55\% of Americans get news from online communities such as Facebook, Twitter, and \reddit~\cite{shearer2019pew}. The scale and speed with which new content is submitted to social media platforms are two key factors that increase the difficulty of how to respond to the spread of misinformation or deceptive news content online, and the appeal of automated or semi-automated defenses or interventions. 

Natural language processing (NLP) models that identify deceptive content offer a path towards fortifying online communities, and a significant body of work (\sect~\ref{sec:related}) has produced countless such models for deception detection tasks~\cite{rubin2016fake, mitra2017parsimonious,  volkova2017separating, rashkin2017truth, karadzhov2018we, shu2020hierarchical}.  
However, evaluation of model performance is typically done in aggregate, across multiple communities, using traditional performance measurements like micro and macro F1-scores.   
We argue that it is critical to understand model behavior at a finer granularity, and  
we evaluate nuanced behavior and failure in the context of the populations that may be affected by predictive outcomes.  

In this work, we seek to {characterize and explain deception detection model performance and biases using the \emph{context} of social media posts---\emph{who} posted the content and \emph{what} community it was posted to.}
{To do so, we compute hundreds of \textit{community} and \textit{author} characteristics using information from two fact checking sources.}

For a given post, community characteristics detail \textit{where} a post was submitted to, \eg \textit{How many links to satirical news sources were submitted to the community this post was submitted to?} Author characteristics detail \textit{who} submitted a post, \eg \textit{How many links to satirical news sources has the author recently submitted?}
Our nuanced evaluation leverages these author and community characteristics to highlight differences in behavior within varying communities or sub-populations, 
to determine whether the model is reliable in general, or if model failures disproportionately impact sub-populations.  

We make use of data from \reddit, a popular social news aggregation platform. Reddit is widely used for research~\cite{medvedev_anatomy_2019} due to its large size and public content~\cite{baumgartner2020pushshift}, and is ideally suited for studying author and community characteristics due to its explicit segmentation into many diverse communities, called ``subreddits'', with different sizes, topics, and user-bases.\footnote{Although our analyses focus exclusively on posts, our approach can easily be extended to include comments in future work. We chose to focus on posts in the current work as they are the primary point of entry for news links submitted to the platform, with many users simply browsing the ranked previews~\cite{glenski2017consumers} as is consistent with social media platforms where a small subset of users typically contribute most new content~\cite{vanmierlo2014rule,hargittai2008participation}.}

%%%%%%%%%%%%%%%%%%%%%%%%%%%%%%%%%%%%%%%%%%%%%
%
% Research Questions
%
%%%%%%%%%%%%%%%%%%%%%%%%%%%%%%%%%%%%%%%%%%%%%%

We use post context (community and author characteristics) and content (text features) to address {two} research questions focused around   
(1) \textit{who} posts deceptive news links and (2) \textit{where} they post differ:
\begin{enumerate}[topsep=3pt,itemsep=3pt]
    \item 
    What characteristics of post authors are associated with high and low model performance? 
    \item  
    How does model performance vary across different communities, and does this correlate with characteristics of those communities? 
\end{enumerate}

{We find that author characteristics are a stronger predictor of high model performance, with the model we evaluate performing especially well on authors who have a history of submitting low factual or deceptive content. We also find that the model performs especially well on posts that are highly accepted by the community, as measured by the community's votes on those posts.}

To our knowledge, we are the first to present a fine-grained evaluation of deception detection model performance in the context of author and community characteristics.

 %%%%%%%%%%%%%%%%%%%%

\section{Related Work}\label{sec:related}

In the last several years, users have seen a tremendous increase in the amount of misinformation, disinformation, and falsified news in circulation on social media platforms.
This seemingly ubiquitous digital deception is in part due to the ease of information dissemination and access on these platforms. 
Many researchers have focused on different areas of detecting deceptive online content. 
\citet{glenski2018humans,kumar2017army,kumar2018rev2} examine the behaviors and activities of malicious users and bots on different social media platforms. 
While others have worked to develop systems to identify fraudulent posts at varying degrees of deception such as broadly classifying suspicious and non-suspicious news \cite{volkova2017separating} to further separating into finer-grained deceptive classes (\eg propaganda, hoax)  
\cite{rashkin2017truth}.

Common amongst recent 
detection methods is the mixed use of machine learning approaches, \eg Random Forest and state-of-the-art deep learning models, \eg Hierarchical Propagation Networks \cite{shu2020hierarchical}.
Of the most prevalent are convolutional neural networks (CNNs)~\cite{ajao2018fake,wang2017liar,volkova2017separating}, Long Short Term Memory (LSTM) neural networks~\cite{ma2016detecting,chen2018unsupervised,rath2017retweet,zubiaga2018discourse,zhang2019reply}, and other variants with attention mechanisms~\cite{guo2018rumor,li2019rumor}.
Designing the right model architecture for a task can be very subjective and laborsome.
Therefore, we implement the binary classification LSTM model from \cite{volkova2019explaining} which reported an F1 score of 0.73 when distinguishing deceptive news from credible.

As artificial intelligence or machine learning models are developed or investigated as potential responses to the issue of misinformation and digital deception online, it is key to understand how models treat the individuals and groups who are impacted by the predictions or recommendations of the models or automated systems. For example, the European Union's GDPR directly addresses the ``right of citizens to receive an explanation for algorithmic decisions''~\cite{goodman2017european} that requires an explanation to be available for individuals impacted by a model decision. Domains outside of deception detection have shown clear evidence of disproportionate biases against certain sub-populations of impacted individuals, \eg predictive policing~\cite{ensign2018runaway}, recidivism prediction~\cite{chouldechova2017fair,dressel2018accuracy}, and hate speech and abusive language identification online~\cite{park2018reducing,davidson2019racial,sap2019risk}. The realm of deception detection is another clear area where disparate performance across communities or certain user groups may have significant negative downstream effects both online and offline. In this work, we seek to go beyond traditional, aggregate performance {metrics} to consider the differing behavior and outcomes of automated deception detection within and across communities and user characteristics.

 %%%%%%%%%%%%%%%%%%%%

\section{Deception Detection Model}\label{sec:task} 
In this work, we focus on a binary classification task to identify posts which link to \textit{Deceptive} or \textit{Credible} news sources.  
{
We evaluate an existing, LSTM-based model architecture previously published by \citet{volkova2019explaining} that relies only on text and lexical features.  
As such, we refer to this model as the ``\LSTMmodelname model.''
}

\subsection{Train and Test Data}\label{sec:train_test_data} 
{To replicate the \LSTMmodelname model for our evaluations, we leverage the previously used list of annotated news sources from \citet{volkova2017separating} as ground truth.
The Volkova annotations consist of two classes: ``Credible\footnote{This class is denoted ``Verified'' in \citet{volkova2017separating}.}'' and ``Deceptive.''
} 
To label individual social media postings linked to these news sources, we propagate annotations of each source to all posts linked to the source. Therefore \textit{Credible posts} are posts which link (via a URL or as posted by the source's official account) to a Credible news source and \textit{Deceptive posts} are posts that link to a news source annotated as Deceptive.

In preliminary experiments, we find that model performance improves when Twitter examples are included in training, even when testing exclusively on \reddit content. A model trained and tested exclusively on \reddit data achieves a test set F1 of 0.577 and we observe a dramatic increase (F1 = 0.725), when we include the Twitter training data. As a result, we focus our analyses using the more robust \LSTMmodelname model trained on both Twitter and \reddit examples.  
As Twitter has no explicit communities equivalent to \reddit subreddits, it is not possible to compute the same community characteristics for Twitter content. As such, in the analyses presented in this paper, we focus exclusively on content posted to \reddit in the test set.

{
To gather train and test data, we collect social media posts from Twitter and \reddit from the same 2016 time period as annotated by \citet{volkova2017separating}. 
} 
For Twitter posts, this resulted in 54.4k Tweets from the official Twitter accounts for news sources that appear in the Volkova annotations. For \reddit content, we collected all link-posts that link to domains associated with the labelled news sources from the Pushshift monthly  archives of \reddit posts\footnote{Pushshift archives of \reddit data were collected from \url{ https://files.pushshift.io/reddit/}}~\cite{baumgartner2020pushshift}, and randomly sample approximately the same number ($\sim54$k) of link-posts as Twitter posts collected.

In order to mitigate the bias of class imbalance on our analyses, these posts were then randomly down-sampled to include an approximately equal number of posts from/linking to deceptive and credible news sources. We divided the resulting data using a random, stratified 80\%/20\% split to create train and test sets, respectively.  

%F1s
%Reddit-Reddit: 0.577
%Reddit-Both: 0.415
%Twitter-Twitter: 0.775
%Twitter-Both: 0.436
%Both-Both: 0.765
%Both-Reddit: 0.725

\section{Community \& Author Characteristics}
{To evaluate fine-grained model performance and biases,}
%To understand the performance of the \LSTMmodelname model, 
we first quantify the \textit{context} in which posts are submitted, using community and author characteristics.

\subsection{Data for Context Annotations}
\label{sec:reddit_data}
{We compute community and author characteristics by examining the entire post history on \reddit for each community and author in the test set.
We use annotations from Volkova \etal (described above, \sect~\ref{sec:train_test_data}) and from Media Bias/Fact Check (MBFC), an independent news source classifier. These annotations were compiled by \citet{weld2021political} and made publicly available\footnote{\url{https://behavioral-data.github.io/news_labeling_reddit/}}. 
}

{
The Volkova \etal annotations provide links to news sources with a categorical label: verified, propaganda, satire, clickbait, conspiracy, and hoax.
The MBFC annotations provide links to news sources with a ordinal label for the factualness of the news source (very low, low, mixed, mostly, high, very high) as well as the political bias (extreme left, left, center left, center, center right, right, extreme right). In addition, the MBFC also include a few categorical labels applicable to a subset of news sources: questionable, satire, conspiracy.
}

 %%%%%%%%%%%%%%%%%%%%

\subsection{Data Validation}\label{sec:validation}
{
Before using these annotations to compute community and author characteristics, we would like to validate that they represent meaningful and accurate aspects of communities and authors, respectively, and are not strongly influenced by noise in the annotation sources.
}
To do so, we assess the coverage of our context annotations,  
--- \ie the fraction of potential news links that we were able to label.

In order to consider the coverage relative to the \textit{potential} news links, we identify a set of domains for which links are definitively not news sources. We identified these non-news links by examining the top 1,000 most frequently linked-to domains across all of reddit and iteratively classified them as non-news based on their domain (\eg reddit-based content hosting domains such as \texttt{v.redd.it} and \texttt{i.redd.it}, external content hosts such as \texttt{imgur.com}, social sites such as  \texttt{facebook.com} and \texttt{instagram.com}, search engines, shopping platforms, music platforms, \etc). 
Websites which were not in English, were not clearly non-news domains, or which did not fit into a clear category, were included in the set of potential news sources. 
We imposed these restrictions to mitigate potential downward bias  
from over-estimating non-news links. 
Although we do not claim to have an exhaustive coverage of non-news links, non-news links included in the set of potential news links at best underrepresents the coverage which is preferable to overrepresentation.

Encouragingly, coverage for both are fairly stable over time, suggesting that there are no significant influxes of additional, unlabelled news sources (or disappearances of retired news sources) that might be biasing our approach. As the MBFC set contains more news sources, the coverage is greater ($\sim 18\%$ on average)  
than the Volkova set ($\sim 10\%$).

\subsection{Community and Author Characteristics}
Using the author and community history collection of posts and the associated MBFC and Volkova \etal annotations, we compute context characteristics for each subreddit community and author that is present in the test set described in \sect~\ref{sec:task}. 

{
First, we compute the general activity of each community and author. These characteristics include the total number of posts by each community or author, the total number of removed posts, and similar overall counts that do not consider the nature of the content submitted.
}

{
Second, for each of the MBFC and Volkova \etal labels (\eg `Satire' from Volkova \etal or `Right Bias' from MBFC) we compute absolute and normalized counts of links of each category for each community and author. Normalized counts for each category are computed by dividing the number of links in the category submitted to each subreddit or by each author by the total number of links submitted in any category. 
This gives, for example, the fraction of links submitted by a author to MBFC High Factual news sources.
}

{
Third, for communities, we compute the equality of contributor activity (number of links submitted per contributor) using the Gini coefficient.
A community with a Gini coefficient  close to 1 would indicate almost all links in that community were submitted by a small fraction of users. On the other hand, a  coefficient close to 0 would indicate that all users of the community who submit links submit approximately the same number of links each.
}

{
Last, again for communities, we approximate the community acceptance by normalizing the score (upvotes - downvotes) of each post relative to the median score of all posts submitted to the subreddit.  
A post with a normalized score of 1 received a typical score for the community it was submitted to, whereas a post with a normalized score of 100 received $100\times$ as many upvotes as a typical post and was more widely or strongly positively received by the community.
}

{
Each of the community characteristics are computed separately for each month, maximizing temporal detail. However, as the typical \reddit user submits far less content each month than the typical subreddit receives, most users' counts for specific link types (\eg MBFC Satire) for any individual month will be 0. To reduce sparsity in the data, we use a rolling sum of all posts submitted by the author in the specified month and the five preceding months to compute author characteristics. 
}

 %%%%%%%%%%%%%%%%%%%%

\section{Evaluation Methodology}
\label{sec:analyses} 

Before our evaluation of model performance across different community or author characteristics and settings, we examine the overall performance of the model on aggregate, using macro F1 score, and the variance of performance within communities. A model with strong aggregate performance may have significant variability within sub-communities, especially those which are underrepresented. We also consider the variability of individual predictive outcomes, such as the confidence of predictions, across each class (deceptive and credible news) to examine the differences in model behavior across classes overall. 
We aim to discover if the model treats all posts, communities, and authors equally, or if there are differences in performance for certain groups that would bias the negative impacts of model error.

\subsection{Comparison to Baselines}\label{sec:baseline_methods}
Next, we frame the performance of the \LSTMmodelname model that classifies posts based on text and linguistic signals relative to naive baselines that randomly classify posts or classify posts based on the typical behavior of authors or communities. 
To this end, we consider three baseline models.

    The \textbf{ Author History Baseline} %This baseline 
    considers the author's history over the previous 6 months (as was used to calculate author characteristics) and computes the fraction of their links to news sources which are deceptive, as defined by the Volkova \etal annotations. It then predicts if a new submission is deceptive or credible with a biased random coin flip, with a probability of predicting deceptive equal to the author's recent tendency to submit deceptive news links (\ie the fraction of news links submitted by the author in the last six months that were linked to deceptive sources).

    The \textbf{Community History Baseline}  
    is similar except that it considers the \textit{community's} tendency to receive deceptive news. This baseline predicts `deceptive' with a probability equal to the fraction of news links submitted to a given subreddit in the last month that were linked to deceptive sources.

    The \textbf{50/50 Baseline}  
    predicts credible/deceptive with an unbiased 50/50 coinflip. No consideration is placed on the content, community, or author.

{
We compare the performance of these baselines with that of the \LSTMmodelname model, providing a reference for its performance as well as an indication of the degree to which community and author characteristics alone are predictive of deceptive content.
}

%DOUBLE COLUMN VERSION
\begin{figure*}[t!]
    \centering 
    \includegraphics[width=1.8\columnwidth]{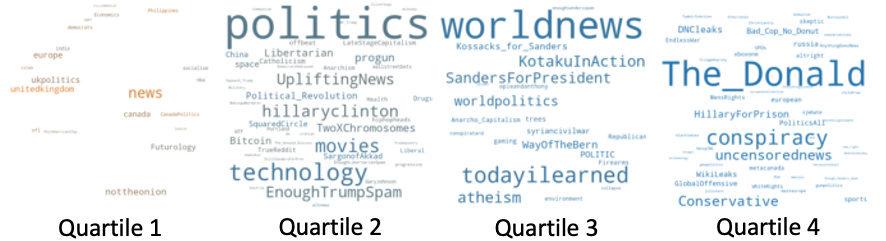}
    \includegraphics[height=1.4in]{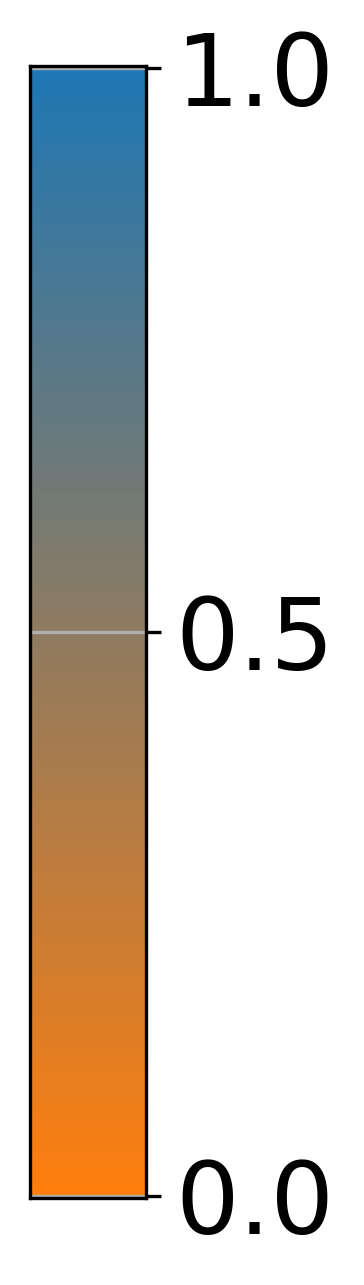} 
    \caption{Communities within each F1 score quartile, represented as wordclouds (size of the community name indicates its volume in test set and the color indicates fine-grained model performance using F1 score).  
    } 
    \label{fig:f1_wordclouds}
\end{figure*}

\subsection{Community and Author Context } 
To better understand how community and author characteristics explain model performance, we compute the Pearson correlation between the value of each characteristic, and the model's confidence in predicting the true class for each post. We compute these correlations across the all test posts, and across deceptive and credible posts (based on true class value) separately. We also examine factors that explain the model's performance on entire authors or communities. To do so, we compute similar correlations for author and community characteristics with aggregated author or community F1 scores, respectively.

\subsection{Popularity and Community Acceptance } 
{
We also examine the relationship between a community's acceptance of a post, and model performance. We measure community acceptance by normalizing each post's score (\# upvotes - \# downvotes) by the median score of a post in that community for the month of submission, to control for the larger number of votes in larger communities.
}
We then compute Pearson's correlations between normalized score and the \LSTMmodelname model's confidence that a post belongs to its annotated class -- here we use not the models prediction confidence but the confidence for the "true class" given the groundtruth labels. 
As before, we use Pearson correlations and a significance threshold of 0.05.

 %%%%%%%%%%%%%%%%%%%%

\section{Results}\label{sec:results}
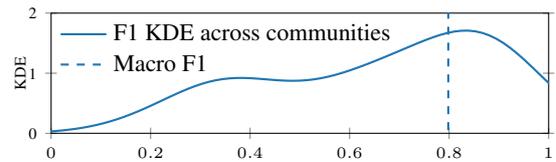
\begin{figure}[t!]
    \centering
    % Read data from csv into \data:
\pgfplotstableread[col sep=comma]{figs/data/model_comparison_data.csv}\data 
 
\small 
 
\begin{tikzpicture} 
\begin{axis}[ 
    %title= F1, % specify title, comment out for no title
    %title style = {yshift=-0.75\baselineskip, font=\small}, 
    tick align=center,
    xtick pos=left,
    ytick={0,1,2,3},
    yticklabels={\textcolor{white}{0.}0, 1,2,3},
    %xlabel=F1,% Specify Label for X-Axis, comment out for no label
    ylabel=KDE,% Specify Label for X-Axis, comment out for no label
    ylabel style={yshift=-\baselineskip},
    height=1.25in, %1.25in, % height of plot
    width=3.2in,  % width of plot
    xmin=0,xmax=1,% x-axis limits
    ymin=0,ymax=2,% y-axis limits
    xtick= {0,0.2,0.4,0.6,0.8,1},% xticks (can specify the yticks similarly) 
    scaled x ticks=false, % don't scale ticks
    scaled y ticks=false, % don't scale ticks
    %every y tick label/.append style={font=\tiny} % makes all yticks be "tiny" fontsize
    legend cell align =left,
    legend style={fill=none, at={(0,1)}, %{(0.5,-0.5)}, 
	anchor=north west,legend columns=1,
	column sep=.01cm,draw=none}, 
    %legend style={fill=none, at={(0.5,-0.5)}, 
	%anchor=north,legend columns=-1,
	%column sep=.2cm,draw=none}, 
	font=\tiny,
    ]
      
    % LSTM across communities
	 \addplot+[tab_blue,no marks, solid,  thick] 
	 table [col sep=comma,x=F1_lstm_x, y=F1_lstm_y]\data;

    % LSTM Macro F1
    \addplot+[tab_blue, no marks, dashed, thick] coordinates{ (0.798189182749583,-100) 
    (0.798189182749583,100)
    };

	 %\small 
\addlegendentry{\small F1 KDE across communities};
\addlegendentry{\small Macro F1};
\end{axis}
 \end{tikzpicture} 
 \vspace{-0.5\baselineskip} 
    \vspace{0.1\baselineskip} 
    \caption{Macro F1 and the kernel density estimation (KDE) of F1 score over communities.  
    While overall deception detection model performance is high, there is significant variability in performance across different online communities.    
    }
    \label{fig:f1_distrib}
\end{figure}

Although the \LSTMmodelname model achieves an overall F1 score on the test set of 0.79, we see that the model performs much better on content from some communities than others (see Figure~\ref{fig:f1_distrib}). Figure~\ref{fig:f1_wordclouds} presents the communities within the test set, partitioned by levels of model performance using the quartiles for the F1 scores. We find that 20\% of the communities {represented} in our test set have F1 $< 0.40$, despite an overall test set F1 of almost 0.8.

In the following subsections, we examine how the model's performance can be explained by community and author characteristics, post popularity, and community acceptance %, and the language used in the post when submitted 
as we seek to understand \textit{why} the model performs far better on content from some communities than others.

\subsection{Comparison to Baselines}\label{sec:baselines}

% MANN-WHITNEY U TEST FOR SIGNIFICANCE THAT AUTHOR BASELINE PERFORMANCE IS GREATER THAN COMMUNITY BASELINE PERFORMANCE

% METRIC          TEST-STAT     P VAL     USR MEAN   SUB MEAN
% precision        +348.00    3.7e-04      0.566      0.714 
% recall           +271.50    1.1e-05      0.042      0.119 
% f1               +477.00    2.7e-02      0.683      0.735 

{We use the community and author history baselines, as well as the 50/50 baseline described in \sect\ref{sec:baseline_methods} to contextualize the performance of the \LSTMmodelname model.}
Figure~\ref{fig:baseline_comparison} presents the distributions of performance across communities for each metric (solid lines) and the overall performance of each model (indicated by the dashed, vertical lines) using three traditional performance metrics: precision, recall, and F1 score. As expected, the \LSTMmodelname model (in blue) dramatically outperforms the 50/50 baseline (in red) on all metrics, and achieves the best performance overall for F1 score (a significant difference in performance, p-value $\leq 1.5\times 10^{-4}$). 

However, the community and author history baselines have very high precision, offset by very poor recall.  In comparing the two, the author baseline significantly outperforms the community baseline on precision, recall, and F1 (p-value $< .02$). This suggests that an author's previous activity is a better predictor of whether an author will submit deceptive content in the future than a community's previous behavior is of whether deceptive content will be submitted to the community in the future. This may be a result of a greater consistency in the behavior of an individual compared to a community where membership may vary over time, if not community attitudes.

\subsection{Community and Author Context} 
\label{sec:char_corrs}

In our next analyses, we investigate how community and author characteristics {correlate with model confidence. We compute these correlations across the entire test set, as well as for just credible and deceptive posts separately.}
 
We summarize the strongest, significant correlations between community or author context and model confidence in Table~\ref{tab:heatmap_confidence}, using a threshold of at least 0.25 for inclusion.
When we examine the author and community characteristics of posts from all classes, the strongest correlation coefficients are all positive, and suggest moderate correlations with stronger model confidence. 
{{The four strongest correlations from the author characteristics pertain to the author's tendency to submit posts linked to questionable or low factual news sources.  
In contrast, the author's tendency to link to \textit{high} factual content is relatively  correlated ($r=-0.21$) with \textit{weaker} model confidence.} 
It is easier for the model to identify \textit{deceptive} posts submitted by authors who typically submit links to low-quality or deceptive news sources. Similarly, we see moderate correlation between increasing presence of deceptive or low factual news media in the community and model performance. Looking at each class individually, we see the strongest relationships for deceptive posts, with little to no correlation for credible posts.

\begin{figure}
    \centering 
    % Read data from csv into \data:
\pgfplotstableread[col sep=comma]{figs/data/model_comparison_data.csv}\data

%{'precision_lstm': 0.7341661187815034,
% 'precision_author': 0.9671052357338802,
% 'precision_subreddit': 0.799628547937001,
% 'precision_random': 0.5615516708338684,
% 'recall_lstm': 0.8744453145392848,
% 'recall_author': 0.19952029520295197,
% 'recall_subreddit': 0.04151575361907466,
% 'recall_random': 0.4999176837922225,
% 'F1_lstm': 0.798189182749583,
% 'F1_author': 0.33075854962537166,
% 'F1_subreddit': 0.07891407600106194,
% 'F1_random': 0.5289199105333809}

\small
 
\begin{tikzpicture} 
\begin{axis}[ 
    title=  {\small Precision}, % specify title, comment out for no title
    title style = {yshift=-0.75\baselineskip, font=\small}, 
    %xlabel=XLABEL,% Specify Label for X-Axis, comment out for no label
    ylabel=KDE,% Specify Label for X-Axis, comment out for no label
    ylabel style={yshift=-0.5\baselineskip},
    height=1.25in, % 1.25in, % height of plot
    width=3.2in,  % width of plot
    xmin=0,xmax=1,% x-axis limits
    ymin=0,ymax=2,% y-axis limits
    xtick pos=left,
    tick align=center,
    xtick= {0,0.2,0.4,0.6,0.8,1},% xticks (can specify the yticks similarly)
    xticklabels=\empty,% hide xtick labels on top plots
    scaled x ticks=false, % don't scale ticks
    scaled y ticks=false, % don't scale ticks
    %every y tick label/.append style={font=\tiny} % makes all yticks be "tiny" fontsize,
    font=\tiny,
    ]
     
%{'precision_lstm': 0.7341661187815034,
% 'precision_author': 0.9671052357338802,
% 'precision_subreddit': 0.799628547937001,
% 'precision_random': 0.5615516708338684,
    % add vertical lines -- e.g. dashed line at 0.5, replace with real x-values
    % 50/50 baseline 
    \addplot+[tab_red, no marks, dashed, thick] coordinates{
    (0.5615516708338684,-100)  
    (0.5615516708338684,100)
    };
    % Author baseline
    \addplot+[tab_orange, dashed, thick] coordinates{ (0.9671052357338802,-100)   
    (0.9671052357338802,100)
    };
    % Subreddit baseline
    \addplot+[tab_green, dashed, thick] coordinates{ (0.799628547937001,-100) 
    (0.799628547937001,100)
    };
    % LSTM
    \addplot+[tab_blue, dashed, thick] coordinates{ (0.7341661187815034,-100)  
    (0.7341661187815034,100)
    };
    
     % add line of color tab_red, using x=XValues, y=YColumnName (e.g.  precision5050, precisionAuthor, precisionSubreddit, precisionLSTM)columns from the csv file read into \data on line 2 (above)
    % 50/50 baseline 
	 \addplot+[tab_red, no marks, solid,  thick] 
	 table [col sep=comma,x=precision_random_x, y=precision_random_y]\data; 
	 
    % Author baseline
	 \addplot+[tab_orange, no marks, solid, thick] 
	 table [col sep=comma,x=precision_author_x, y=precision_author_y]\data; 
	 
    % Subreddit baseline
	 \addplot+[tab_green,no marks, solid,  thick] 
	 table [col sep=comma,x=precision_subreddit_x, y=precision_subreddit_y]\data; 
	 
    % LSTM
	 \addplot+[tab_blue,no marks, solid,  thick] 
	 table [col sep=comma,x=precision_lstm_x, y=precision_lstm_y]\data; 
\end{axis}
 \end{tikzpicture} \vspace{-0.25\baselineskip}

%\hspace{-0.1\baselineskip}
\begin{tikzpicture} 
\begin{axis}[ 
    title= {\small Recall},% (Log-Scale Density), % specify title, comment out for no title
    ymode = log,
    tick align=center,
    xtick pos=left,
    ytick = {0.1, 1, 10},
    yticklabels = {0.1, 1, 10},
    title style = {yshift=-0.75\baselineskip, font=\small}, 
    %xlabel=XLABEL,% Specify Label for X-Axis, comment out for no label
    ylabel=KDE (log),% Specify Label for X-Axis, comment out for no label
    ylabel style={yshift=-0.52\baselineskip},
    height=1.25in, % 1.25in, % height of plot
    width=3.2in,  % width of plot
    xmin=0,xmax=1,% x-axis limits
    ymin=0.1,ymax=25,% y-axis limits
    xtick= {0,0.2,0.4,0.6,0.8,1},% xticks (can specify the yticks similarly)
    xticklabels=\empty,% hide xtick labels on top plots
    scaled x ticks=false, % don't scale ticks
    scaled y ticks=false, % don't scale ticks
    %every y tick label/.append style={font=\tiny} % makes all yticks be "tiny" fontsize
    font=\tiny,
    ]
     
% 'recall_lstm': 0.8744453145392848,
% 'recall_author': 0.19952029520295197,
% 'recall_subreddit': 0.04151575361907466,
% 'recall_random': 0.4999176837922225,
    % add vertical lines -- e.g. dashed line at 0.5, replace with real x-values
    % 50/50 baseline 
    \addplot+[tab_red, no marks, dashed, thick] coordinates{ (0.4999176837922225,0.1)%-100)
    (0.4999176837922225,100)
    };
    % Author baseline
    \addplot+[tab_orange, no marks, dashed, thick] coordinates{ (0.19952029520295197,0.1)%-100) 
    (0.19952029520295197,100)
    };
    % Subreddit baseline
    \addplot+[tab_green, no marks, dashed, thick] coordinates{ (0.04151575361907466,0.1)%-100)  
    (0.04151575361907466,100)
    };
    % LSTM
    \addplot+[tab_blue, no marks, dashed, thick] coordinates{ (0.8744453145392848,0.1)%-100) 
    (0.8744453145392848,100)
    };
    
     % add line of color tab_red, using x=XValues, y=YColumnName (e.g.  precision5050, precisionAuthor, precisionSubreddit, precisionLSTM)columns from the csv file read into \data on line 2 (above)
    % 50/50 baseline 
	 \addplot+[tab_red, no marks, solid,   thick] 
	 table [col sep=comma,x=recall_random_x, y=recall_random_y]\data; 
	 
    % Author baseline
	 \addplot+[tab_orange, no marks, solid,   thick] 
	 table [col sep=comma,x=recall_author_x, y=recall_author_y]\data; 
	 
    % Subreddit baseline
	 \addplot+[tab_green,no marks,  solid,   thick] 
	 table [col sep=comma,x=recall_subreddit_x, y=recall_subreddit_y]\data; 
	 
    % LSTM
	 \addplot+[tab_blue,no marks, solid,    thick] 
	 table [col sep=comma,x=recall_lstm_x, y=recall_lstm_y]\data; 
\end{axis}
 \end{tikzpicture} \vspace{-0.25\baselineskip}

\begin{tikzpicture} 
\begin{axis}[ 
    title= {\small F1}, % specify title, comment out for no title
    title style = {yshift=-0.75\baselineskip, font=\small}, 
    tick align=center,
    xtick pos=left,
    ytick={0,1,2,3},
    yticklabels={\textcolor{white}{0.}0, 1,2,3},
    %xlabel=XLABEL,% Specify Label for X-Axis, comment out for no label
    ylabel=KDE,% Specify Label for X-Axis, comment out for no label
    ylabel style={yshift=-0.5\baselineskip},
    height=1.25in, % 1.25in, % height of plot
    width=3.2in,  % width of plot
    xmin=0,xmax=1,% x-axis limits
    ymin=0,ymax=3,% y-axis limits
    ytick ={0, 1,2,3},
    yticklabels ={\textcolor{white}{0.}0, 1,2,3},
    xtick= {0,0.2,0.4,0.6,0.8,1},% xticks (can specify the yticks similarly) 
    scaled x ticks=false, % don't scale ticks
    scaled y ticks=false, % don't scale ticks
    %every y tick label/.append style={font=\tiny} % makes all yticks be "tiny" fontsize
    font=\tiny,
    ]
     
% 'F1_lstm': 0.798189182749583,
% 'F1_author': 0.33075854962537166,
% 'F1_subreddit': 0.07891407600106194,
% 'F1_random': 0.5289199105333809}
    % add vertical lines -- e.g. dashed line at 0.5, replace with real x-values
    % 50/50 baseline 
    \addplot+[tab_red, no marks, dashed, thick] coordinates{ (0.5289199105333809,-100) 
    (0.5289199105333809,100)
    };
    % Author baseline
    \addplot+[tab_orange, no marks, dashed, thick] coordinates{ (0.33075854962537166,-100) 
    (0.33075854962537166,100)
    };
    % Subreddit baseline
    \addplot+[tab_green, no marks, dashed, thick] coordinates{ (0.07891407600106194,-100) 
    (0.07891407600106194,100)
    };
    % LSTM
    \addplot+[tab_blue, no marks, dashed, thick] coordinates{ (0.798189182749583,-100) 
    (0.798189182749583,100)
    };
    
     % add line of color tab_red, using x=XValues, y=YColumnName (e.g.  precision5050, precisionAuthor, precisionSubreddit, precisionLSTM)columns from the csv file read into \data on line 2 (above)
    % 50/50 baseline 
	 \addplot+[tab_red, no marks, solid, thick] 
	 table [col sep=comma,x=F1_random_x, y=F1_random_y]\data; 
	 
    % Author baseline
	 \addplot+[tab_orange, no marks, solid,  thick] 
	 table [col sep=comma,x=F1_author_x, y=F1_author_y]\data; 
	 
    % Subreddit baseline
	 \addplot+[tab_green,no marks, solid,  thick] 
	 table [col sep=comma,x=F1_subreddit_x, y=F1_subreddit_y]\data; 
	 
    % LSTM
	 \addplot+[tab_blue,no marks, solid,  thick] 
	 table [col sep=comma,x=F1_lstm_x, y=F1_lstm_y]\data; 
\end{axis}
 \end{tikzpicture} \vspace{-0.25\baselineskip}
    \begin{tikzpicture} 
\begin{axis}[%
    hide axis, height =.75in,
    xmin=0,xmax=50,ymin=0,ymax=0.4,
	legend cell align=left,
    legend style={fill=none, at={(0.5,1)},%0.5)}, 
	    anchor=north,legend columns=2,column sep=.15cm,draw=none},   
]
\addlegendimage{no marks, tab_blue ,thick}
\addlegendentry{\LSTMmodelname};

\addlegendimage{no marks, tab_orange ,thick}
\addlegendentry{Author History Baseline};

\addlegendimage{no marks, tab_red ,thick}
\addlegendentry{50/50 Baseline};

\addlegendimage{no marks, tab_green ,thick}
\addlegendentry{Community History Baseline};

\end{axis}
\end{tikzpicture}
    \vspace{-0.5\baselineskip}
    \caption{KDE plots illustrating performance metric distributions across communities for the \LSTMmodelname and baseline models. Vertical dashed lines indicate the aggregate metric for each model across the full test set. While the \LSTMmodelname model achieves the best overall performance and recall, the author and community characteristic baselines have higher precision.}
    \label{fig:baseline_comparison}
    
% MANN-WHITNEY U TEST FOR SIGNIFICANCE THAT BASELINE PERFORMANCE IS LESS THAN LSTM PERFORMANCE

% BASELINE        TEST-STAT     P VAL     BASELINE MEAN    LSTM MEAN
% PRECISION
% sub              +559.00    1.6e-01      0.566             0.519 
% usr              +928.50    1.0e+00      0.714             0.519 
% random           +497.50    4.6e-02      0.361             0.519 

% RECALL
% sub               +68.50    3.4e-11      0.042             0.783 
% usr               +89.00    1.6e-10      0.119             0.783 
% random           +229.00    1.2e-06      0.500             0.783 

% F1
% sub              +249.00    3.6e-06      0.683             0.853 
% usr              +249.00    3.6e-06      0.735             0.853 
% random           +327.00    1.5e-04      0.773             0.853 

\end{figure}
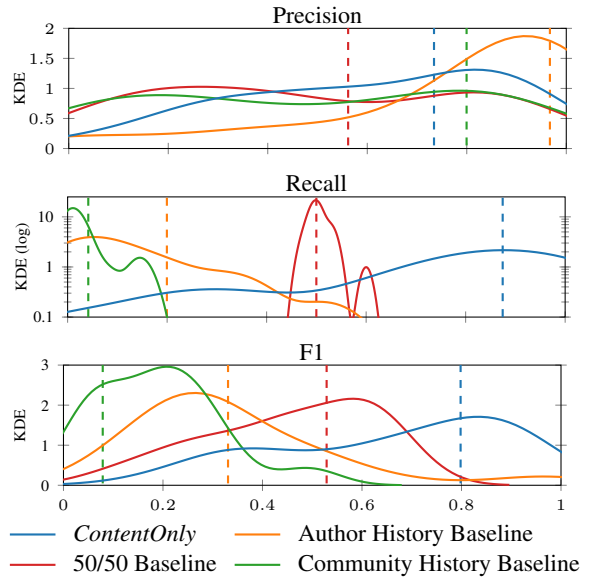

\begin{table}[t!]
    \small  
    \centering
    
\newcommand{\thresholdTwentyFive}[1]{}

%\small
\footnotesize
 \vspace{0.5\baselineskip}
 \setlength\tabcolsep{3 pt} %default is 6
\begin{tabular}{@{\hskip .025cm}cr|lll}
%{@{\hskip .05cm}cr|lll}
\toprule
%{} & &     \multicolumn{1}{c}{\footnotesize \textit{All}} &                                     \multicolumn{1}{c}{\footnotesize \textit{Trusted}} &   \multicolumn{1}{c}{\footnotesize \textit{Deceptive}} \\

{} & &     \multicolumn{1}{c}{  \textit{All}} &                                     \multicolumn{1}{c}{ \textit{C Posts}} &   \multicolumn{1}{c}{  \textit{D Posts}} \\
\midrule \multirow{7}{*}{\rot{\bf \small \textit{Author's Links} }} & \multicolumn{1}{l|}{ \textit{  Bias  } } \\
%%\midrule \multirow{10}{*}{\rot{\bf \it \footnotesize Author's Characteristics}} & \multicolumn{1}{l|}{ \textit{ \footnotesize Bias  } } \\

{} & \% Center Right Bias & \cellcolor[rgb]{0.860,0.942,0.747} \bf +0.25$\ddag$ & \cellcolor[rgb]{0.969,0.967,0.968} -0.00 & \cellcolor[rgb]{0.831,0.930,0.702} \bf +0.27$\ddag$ \\
%\cline{2-5} {} 
& \multicolumn{1}{l|}{ \textit{   Categorical  } } \\

\thresholdTwentyFive{

{} & \% Unlabeled (Volkova) 
& \cellcolor[rgb]{0.983,0.914,0.950} -0.12$\ddag$ & \cellcolor[rgb]{0.977,0.935,0.957} -0.07$\ddag$ & \cellcolor[rgb]{0.988,0.866,0.931} \bf -0.21$\ddag$ \\
{} & \% Trusted (Volkova) 
& \cellcolor[rgb]{0.988,0.866,0.931} \bf -0.21$\ddag$ & \cellcolor[rgb]{0.949,0.966,0.924} +0.06$\ddag$ & \cellcolor[rgb]{0.949,0.966,0.924} +0.06$\ddag$ \\

}

{} & \textcolor{white}{....} \% Propaganda (Volkova) & \cellcolor[rgb]{0.760,0.899,0.588} \bf +0.35$\ddag$ & \cellcolor[rgb]{0.941,0.965,0.906} +0.08$\ddag$ & \cellcolor[rgb]{0.895,0.958,0.804}  +0.20$\ddag$ \\
{} & \% Questionable (MBFC) & \cellcolor[rgb]{0.753,0.896,0.577} \bf +0.37$\ddag$ & \cellcolor[rgb]{0.936,0.965,0.894} +0.10$\ddag$ & \cellcolor[rgb]{0.895,0.958,0.804}  +0.20$\ddag$ \\
%\cline{2-5} {} 
& \multicolumn{1}{l|}{ \textit{   Factualness  } } \\
{} & \% Very Low Factual & \cellcolor[rgb]{0.782,0.908,0.622} \bf +0.33$\ddag$ & \cellcolor[rgb]{0.946,0.966,0.918} +0.07$\ddag$ & \cellcolor[rgb]{0.824,0.927,0.691} \bf +0.29$\ddag$ \\

\thresholdTwentyFive{

{} & \% High Factual & \cellcolor[rgb]{0.988,0.866,0.931} \bf -0.21$\ddag$ & \cellcolor[rgb]{0.952,0.967,0.930} +0.05$\ddag$ & \cellcolor[rgb]{0.974,0.949,0.962} -0.04$\dag$ \\

}

\midrule \multirow{10}{*}{\rot{\bf \small \textit{Community's Links}}} & \multicolumn{1}{l|}{ \textit{   Bias  } } \\
%\midrule \multirow{36}{*}{\rot{\bf \it \footnotesize Community's Characteristics}} & \multicolumn{1}{l|}{ \textit{ \footnotesize Bias  } } \\

%{} & \# Center Bias Links & \cellcolor[rgb]{0.910,0.962,0.834} +0.17$\ddag$ & \cellcolor[rgb]{0.912,0.962,0.840} +0.17$\ddag$ & \cellcolor[rgb]{0.902,0.961,0.816}  +0.20$\ddag$ \\
{} & \# Right Bias & \cellcolor[rgb]{0.860,0.942,0.747} \bf +0.25$\ddag$ & \cellcolor[rgb]{0.907,0.961,0.828} +0.18$\ddag$ & \cellcolor[rgb]{0.941,0.965,0.906} +0.08$\ddag$ \\

\thresholdTwentyFive{

{} & \# Extreme Right Bias & \cellcolor[rgb]{0.874,0.948,0.770} \bf +0.23$\ddag$ & \cellcolor[rgb]{0.918,0.963,0.852} +0.15$\ddag$ & \cellcolor[rgb]{0.954,0.967,0.936} +0.04$\dag$ \\

{} & \# Unlabeled Bias & \cellcolor[rgb]{0.895,0.958,0.804} \bf +0.21$\ddag$ & \cellcolor[rgb]{0.944,0.966,0.912} +0.08$\ddag$ & \cellcolor[rgb]{0.941,0.965,0.906} +0.08$\ddag$ \\

}

%\cline{2-5} {} 
& \multicolumn{1}{l|}{ \textit{   Categorical  } } \\

\thresholdTwentyFive{

{} & \# Retired (MBFC) & \cellcolor[rgb]{0.881,0.952,0.782} \bf +0.23$\ddag$ & \cellcolor[rgb]{0.910,0.962,0.834} +0.17$\ddag$ & \cellcolor[rgb]{0.954,0.967,0.936} +0.05$\ddag$ \\

}

{} & \# Clickbait (Volkova) & \cellcolor[rgb]{0.860,0.942,0.747} \bf +0.25$\ddag$ & \cellcolor[rgb]{0.912,0.962,0.840} +0.17$\ddag$ & \cellcolor[rgb]{0.949,0.966,0.924} +0.06$\ddag$ \\

\thresholdTwentyFive{

{} & \# Hoax (Volkova) & \cellcolor[rgb]{0.888,0.955,0.793} \bf +0.22$\ddag$ & \cellcolor[rgb]{0.931,0.964,0.882} +0.12$\ddag$ & \cellcolor[rgb]{0.944,0.966,0.912} +0.08$\ddag$ \\
{} & \# Propaganda (Volkova) & \cellcolor[rgb]{0.874,0.948,0.770} \bf +0.23$\ddag$ & \cellcolor[rgb]{0.925,0.964,0.870} +0.13$\ddag$ & \cellcolor[rgb]{0.931,0.964,0.882} +0.11$\ddag$ \\
{} & \# Conspiracy (MBFC) & \cellcolor[rgb]{0.874,0.948,0.770} \bf +0.23$\ddag$ & \cellcolor[rgb]{0.920,0.963,0.858} +0.14$\ddag$ & \cellcolor[rgb]{0.949,0.966,0.924} +0.05$\ddag$ \\

}

{} & \# Questionable (MBFC) & \cellcolor[rgb]{0.845,0.936,0.725} \bf +0.26$\ddag$ & \cellcolor[rgb]{0.910,0.962,0.834} +0.18$\ddag$ & \cellcolor[rgb]{0.944,0.966,0.912} +0.08$\ddag$ \\

\thresholdTwentyFive{

{} & \% Questionable (MBFC) & \cellcolor[rgb]{0.881,0.952,0.782} \bf +0.23$\ddag$ & \cellcolor[rgb]{0.902,0.961,0.816} +0.20$\ddag$ & \cellcolor[rgb]{0.967,0.968,0.966} +0.00 \\

{} & \# Unlabeled (Volkova) & \cellcolor[rgb]{0.888,0.955,0.793} \bf +0.21$\ddag$ & \cellcolor[rgb]{0.936,0.965,0.894} +0.10$\ddag$ & \cellcolor[rgb]{0.941,0.965,0.906} +0.08$\ddag$ \\
{} & \# Unlabeled (MBFC) & \cellcolor[rgb]{0.895,0.958,0.804} \bf +0.21$\ddag$ & \cellcolor[rgb]{0.944,0.966,0.912} +0.08$\ddag$ & \cellcolor[rgb]{0.941,0.965,0.906} +0.08$\ddag$ \\

}

%\cline{2-5} {} 
& \multicolumn{1}{l|}{ \textit{  Factualness  } } \\
{} & \# Very Low Factual & \cellcolor[rgb]{0.803,0.918,0.656} \bf +0.31$\ddag$ & \cellcolor[rgb]{0.918,0.963,0.852} +0.15$\ddag$ & \cellcolor[rgb]{0.910,0.962,0.834} +0.17$\ddag$ \\

\thresholdTwentyFive{

{} & \% Very Low Factual & \cellcolor[rgb]{0.895,0.958,0.804} \bf +0.21$\ddag$ & \cellcolor[rgb]{0.920,0.963,0.858} +0.14$\ddag$ & \cellcolor[rgb]{0.928,0.964,0.876} +0.12$\ddag$ \\

}

{} & \# Low Factual & \cellcolor[rgb]{0.860,0.942,0.747} \bf +0.25$\ddag$ & \cellcolor[rgb]{0.915,0.962,0.846} +0.16$\ddag$ & \cellcolor[rgb]{0.944,0.966,0.912} +0.07$\ddag$ \\

\thresholdTwentyFive{

{} & \# Mixed Factual & \cellcolor[rgb]{0.867,0.945,0.759} \bf +0.24$\ddag$ & \cellcolor[rgb]{0.907,0.961,0.828} +0.18$\ddag$ & \cellcolor[rgb]{0.939,0.965,0.900} +0.09$\ddag$ \\

{} & \% Mixed Factual & \cellcolor[rgb]{0.907,0.961,0.828} +0.18$\ddag$ & \cellcolor[rgb]{0.895,0.958,0.804} \bf +0.21$\ddag$ & \cellcolor[rgb]{0.971,0.960,0.966} -0.02 \\

{} & \# Very High Factual & \cellcolor[rgb]{0.918,0.963,0.852} +0.15$\ddag$ & \cellcolor[rgb]{0.923,0.963,0.864} +0.14$\ddag$ & \cellcolor[rgb]{0.867,0.945,0.759} \bf +0.24$\ddag$ \\

{} & \# Unlabeled Factualness & \cellcolor[rgb]{0.895,0.958,0.804} \bf +0.21$\ddag$ & \cellcolor[rgb]{0.944,0.966,0.912} +0.08$\ddag$ & \cellcolor[rgb]{0.941,0.965,0.906} +0.08$\ddag$ \\

}

%\cline{2-5}  {} 

\thresholdTwentyFive{

& \multicolumn{1}{l|}{ \textit{  General  } } \\ 
{} & \# Posts & \cellcolor[rgb]{0.881,0.952,0.782} \bf +0.22$\ddag$ & \cellcolor[rgb]{0.928,0.964,0.876} +0.12$\ddag$ & \cellcolor[rgb]{0.939,0.965,0.900} +0.09$\ddag$ \\
%{} & \# Selfposts & \cellcolor[rgb]{0.895,0.958,0.804} +0.20$\ddag$ & \cellcolor[rgb]{0.944,0.966,0.912} +0.07$\ddag$ & \cellcolor[rgb]{0.949,0.966,0.924} +0.06$\ddag$ \\
{} & \# Linkposts & \cellcolor[rgb]{0.881,0.952,0.782} \bf +0.22$\ddag$ & \cellcolor[rgb]{0.931,0.964,0.882} +0.12$\ddag$ & \cellcolor[rgb]{0.933,0.964,0.888} +0.10$\ddag$ \\
{} & \# Deleted Posts & \cellcolor[rgb]{0.888,0.955,0.793} \bf +0.22$\ddag$ & \cellcolor[rgb]{0.925,0.964,0.870} +0.13$\ddag$ & \cellcolor[rgb]{0.944,0.966,0.912} +0.07$\ddag$ \\
{} & \# Links & \cellcolor[rgb]{0.888,0.955,0.793} \bf +0.22$\ddag$ & \cellcolor[rgb]{0.928,0.964,0.876} +0.12$\ddag$ & \cellcolor[rgb]{0.939,0.965,0.900} +0.09$\ddag$ \\
{} & \# Twitter Links & \cellcolor[rgb]{0.867,0.945,0.759} \bf +0.24$\ddag$ & \cellcolor[rgb]{0.931,0.964,0.882} +0.11$\ddag$ & \cellcolor[rgb]{0.933,0.964,0.888} +0.10$\ddag$ \\
{} & \# Potential News Links & \cellcolor[rgb]{0.888,0.955,0.793} \bf +0.21$\ddag$ & \cellcolor[rgb]{0.928,0.964,0.876} +0.12$\ddag$ & \cellcolor[rgb]{0.918,0.963,0.852} +0.15$\ddag$ \\
{} & \# Non-News Social Links & \cellcolor[rgb]{0.888,0.955,0.793} \bf +0.22$\ddag$ & \cellcolor[rgb]{0.936,0.965,0.894} +0.09$\ddag$ & \cellcolor[rgb]{0.944,0.966,0.912} +0.07$\ddag$ \\
%{} & \% Non-News Reference Links & \cellcolor[rgb]{0.988,0.896,0.943} -0.16$\ddag$ & \cellcolor[rgb]{0.992,0.878,0.937}  -0.20$\ddag$ & \cellcolor[rgb]{0.988,0.896,0.943} -0.16$\ddag$ \\
%{} & \# Non-News External Host Links & \cellcolor[rgb]{0.895,0.958,0.804} +0.20$\ddag$ & \cellcolor[rgb]{0.941,0.965,0.906} +0.08$\ddag$ & \cellcolor[rgb]{0.957,0.967,0.942} +0.04$\dag$ \\

}

%\cline{2-5} {} 
& \multicolumn{1}{l|}{ \textit{  Inequality  } } \\
{} & Gini Coefficient (\# Links) & \cellcolor[rgb]{0.796,0.915,0.645} \bf +0.32$\ddag$ & \cellcolor[rgb]{0.910,0.962,0.834} +0.18$\ddag$ & \cellcolor[rgb]{0.925,0.964,0.870} +0.13$\ddag$ \\
\midrule 
%\multicolumn{1}{l}{ \rot{ \it Post  } }
& Post Score & \cellcolor[rgb]{0.928,0.964,0.876} +0.12$\ddag$ & \cellcolor[rgb]{0.978,0.931,0.956} -0.08$\ddag$ & \cellcolor[rgb]{0.845,0.936,0.725} {\bf +0.27$\ddag$} \\ \midrule
\end{tabular}
    \caption{Correlations between community and author characteristics and true class confidence across the entire test set (All), credible posts (C posts), or deceptive posts (D posts). {Characteristics are included when correlation $|r|\ge.25$ (in bold) in at least one column.}   
    $\dag$ denotes a p-value $<.05$,
    $\ddag$ 
    denotes a p-value $<.01$.
    }
    \label{tab:heatmap_confidence}
\end{table}

\begin{table*}[thb!]
\footnotesize %\small
    \centering
\small
\footnotesize
\setlength\tabcolsep{3 pt} %default is 6
\begin{tabular}{p{0in}@{\hskip 0in}lrlr|lrlr}% lp{0.65cm}lp{0.65cm}|lp{0.8cm}lp{0.9cm}}
%{p{0.1cm}|lp{0.4cm}lp{0.4cm}|lp{0.65cm}lp{0.65cm}}
\toprule
 & \multicolumn{4}{l|}{\textbf{Features Positively Correlated with Metric}} & \multicolumn{4}{l}{\textbf{Features Negatively Correlated with Metric}} \\
& \multicolumn{1}{l}{Community Feature} & \multicolumn{1}{r}{$r$} & \multicolumn{1}{l}{Author Feature} & \multicolumn{1}{r|}{$r$}  & \multicolumn{1}{l}{Community Feature} & \multicolumn{1}{r}{$r$} & \multicolumn{1}{l}{Author Feature} & \multicolumn{1}{r}{$r$} \\
\midrule
\multicolumn{9}{l}{\footnotesize \it F1-Score}\\
  \midrule
\multirow[c]{7}{*}{}&%F1~} &

\textcolor{frac_true_deceptive}{\rule{0.2cm}{0.2cm}} %\cellcolor{frac_true_deceptive!15}
  ($\%_{T}$) Deceptive &  %\cellcolor[rgb]{0.299,0.569,0.129}
 \textcolor{black}{0.80\ddag} &

 \textcolor{frac_true_deceptive}{\rule{0.2cm}{0.2cm}} %\cellcolor{frac_true_deceptive!15}  
  ($\%_{T}$) Deceptive   &  %\cellcolor[rgb]{0.264,0.527,0.121}
 \textcolor{black}{0.85\ddag} &  
 
  \textcolor{mean_frac_high_factual}{\rule{0.2cm}{0.2cm}} % \cellcolor{mean_frac_high_factual!15} 
   ($\%_{L}$) High Factual  &  %\cellcolor[rgb]{0.935,0.680,0.831}
  -0.43\ddag &    
  
   \textcolor{mean_frac_mostly_factual}{\rule{0.2cm}{0.2cm}} %\cellcolor{mean_frac_mostly_factual!15}
     ($\%_{L}$) Mostly Factual &   %\cellcolor[rgb]{0.821,0.283,0.584}
   \textcolor{black}{-0.70\dag} \\
  &           
  
   \textcolor{mean_frac_pnnl_deceptive}{\rule{0.2cm}{0.2cm}} %\cellcolor{mean_frac_pnnl_deceptive!15}
   ($\%_{L}$) Deceptive  &  %\cellcolor[rgb]{0.725,0.884,0.531} 
   0.40\ddag &                                 \cellcolor{gray!5} &                 \cellcolor{gray!5} &                              \cellcolor{gray!5} &                  \cellcolor{gray!5} &   
   
    \textcolor{mean_frac_high_factual}{\rule{0.2cm}{0.2cm}} %\cellcolor{mean_frac_high_factual!15} 
    ($\%_{L}$) High Factual  &  %\cellcolor[rgb]{0.938,0.690,0.838}
    -0.42\ddag \\
  &       
  
   \textcolor{mean_frac_low_factual}{\rule{0.2cm}{0.2cm}} %\cellcolor{mean_frac_low_factual!15} 
    ($\%_{L}$) Low Factual  &  %\cellcolor[rgb]{0.725,0.884,0.531}
   0.39\ddag &                                 \cellcolor{gray!5} &                 \cellcolor{gray!5} &                              \cellcolor{gray!5} &                  \cellcolor{gray!5} &     
   
    \textcolor{mean_frac_center_bias}{\rule{0.2cm}{0.2cm}} %\cellcolor{mean_frac_center_bias!15} 
    ($\%_{L}$) Center Bias &   %\cellcolor[rgb]{0.955,0.749,0.873}
    -0.36\dag \\
  &  
  
   \textcolor{mean_frac_very_low_factual}{\rule{0.2cm}{0.2cm}} %\cellcolor{mean_frac_very_low_factual!15}
    ($\%_{L}$) V. Low Factual  &   %\cellcolor[rgb]{0.746,0.893,0.565}
   0.37\dag &                                 \cellcolor{gray!5} &                 \cellcolor{gray!5} &                              \cellcolor{gray!5} &                  \cellcolor{gray!5} &     
   
  \ignore{
    \textcolor{mean_frac_left_bias}{\rule{0.2cm}{0.2cm}} %\cellcolor{mean_frac_left_bias!15} 
     ($\%_{L}$)  Left Bias &   %\cellcolor[rgb]{0.966,0.788,0.892}
    -0.30\dag \\
 & 
 }
 \cellcolor{gray!5} &                  \cellcolor{gray!5} \\
  &  
 
  \textcolor{mean_frac_extreme_right_bias}{\rule{0.2cm}{0.2cm}} %\cellcolor{mean_frac_extreme_right_bias!15} 
  ($\%_{L}$) Extr. Right Bias &   %\cellcolor[rgb]{0.760,0.899,0.588} 
  0.36\dag &                                 \cellcolor{gray!5} &                 \cellcolor{gray!5} &                              \cellcolor{gray!5} &                  \cellcolor{gray!5} &                                   \cellcolor{gray!5} &                  \cellcolor{gray!5} \\
  &  
  
   \textcolor{mean_frac_mixed_factual}{\rule{0.2cm}{0.2cm}} %\cellcolor{mean_frac_mixed_factual!15}  
   ($\%_{L}$) Mixed Factual  &   %\cellcolor[rgb]{0.782,0.908,0.622}
   0.33\dag &                                 \cellcolor{gray!5} &                 \cellcolor{gray!5} &                              \cellcolor{gray!5} &                  \cellcolor{gray!5} &                                   \cellcolor{gray!5}  &                  \cellcolor{gray!5}  \\

   \ignore{ % remove lowest & only #
  &       
  
   \textcolor{mean_num_pnnl_deceptive}{\rule{0.2cm}{0.2cm}} % \cellcolor{mean_num_pnnl_deceptive!15} 
   ($\#_{L}$)  Deceptive &   %\cellcolor[rgb]{0.810,0.921,0.668}
   0.30\dag &                                 \cellcolor{gray!5} &                 \cellcolor{gray!5} &                              \cellcolor{gray!5} &                  \cellcolor{gray!5} &                                   \cellcolor{gray!5} &                  \cellcolor{gray!5}  
   \\ 
%&\multicolumn{8}{c}{\cellcolor{gray!5} \it Metric = F1 Score}\\ 
   }

  \midrule
\multicolumn{9}{l}{\footnotesize \it Precision}\\
  \midrule
\multirow[c]{7}{*}{}&% P} &   

 \textcolor{frac_true_deceptive}{\rule{0.2cm}{0.2cm}}  %\cellcolor{frac_true_deceptive!15} 
  ($\%_{T}$) Deceptive &  %\cellcolor[rgb]{0.235,0.491,0.115} 
 \textcolor{black}{0.89\ddag} &
 
  \textcolor{frac_true_deceptive}{\rule{0.2cm}{0.2cm}}  %\cellcolor{frac_true_deceptive!15}
  ($\%_{T}$) Deceptive  &  %\cellcolor[rgb]{0.229,0.484,0.114} 
  \textcolor{black}{0.89\ddag} & 
  
   \textcolor{mean_frac_high_factual}{\rule{0.2cm}{0.2cm}} % \cellcolor{mean_frac_high_factual!15}
   ($\%_{L}$) High Factual  &   %\cellcolor[rgb]{0.963,0.775,0.886}
   -0.33\dag & 
   
    \textcolor{mean_frac_very_high_factual}{\rule{0.2cm}{0.2cm}}  %\cellcolor{mean_frac_very_high_factual!15}
     ($\%_{L}$) V. High Factual  &   %\cellcolor[rgb]{0.914,0.612,0.784}
    -0.48\dag \\
 &    
 
  \textcolor{mean_frac_low_factual}{\rule{0.2cm}{0.2cm}} %\cellcolor{mean_frac_low_factual!15} 
  ($\%_{L}$) Low Factual  &  %\cellcolor[rgb]{0.621,0.817,0.403}
  0.48\ddag &     
  
  \textcolor{mean_frac_pnnl_deceptive}{\rule{0.2cm}{0.2cm}} %\cellcolor{mean_frac_pnnl_deceptive!15} 
  ($\%_{L}$) Deceptive  &  %\cellcolor[rgb]{0.782,0.908,0.622} 
  0.33\ddag &                              \cellcolor{gray!5} &                  \cellcolor{gray!5} &  
  
   \textcolor{mean_frac_high_factual}{\rule{0.2cm}{0.2cm}} % \cellcolor{mean_frac_high_factual!15} 
   ($\%_{L}$) High Factual &   %\cellcolor[rgb]{0.923,0.641,0.804} 
   -0.46\dag \\
 & 
 
  \textcolor{mean_frac_extreme_right_bias}{\rule{0.2cm}{0.2cm}} % \cellcolor{mean_frac_extreme_right_bias!15} 
   ($\%_{L}$) Extr. Right Bias &  %\cellcolor[rgb]{0.656,0.840,0.446} 
  0.45\ddag &                                 \cellcolor{gray!5} &                 \cellcolor{gray!5} &                              \cellcolor{gray!5} &                  \cellcolor{gray!5} &  
  
   \textcolor{mean_frac_center_left_bias}{\rule{0.2cm}{0.2cm}} %\cellcolor{mean_frac_center_left_bias!15}  
   ($\%_{L}$) Center Left Bias &   %\cellcolor[rgb]{0.926,0.651,0.811}
   -0.45\dag \\
 &    
 
  \textcolor{mean_frac_mixed_factual}{\rule{0.2cm}{0.2cm}} % \cellcolor{mean_frac_mixed_factual!15}
  ($\%_{L}$) Mixed Factual &  %\cellcolor[rgb]{0.682,0.857,0.478} 
  0.44\ddag &                                 \cellcolor{gray!5} &                 \cellcolor{gray!5} &                              \cellcolor{gray!5} &                  \cellcolor{gray!5} &  
  
    \textcolor{mean_frac_center_bias}{\rule{0.2cm}{0.2cm}}  %\cellcolor{mean_frac_center_bias!15}
    ($\%_{L}$) Center Bias &   %\cellcolor[rgb]{0.944,0.709,0.852} 
  -0.40\dag \\
 &  
 
  \textcolor{mean_frac_very_low_factual}{\rule{0.2cm}{0.2cm}} %\cellcolor{mean_frac_very_low_factual!15}  
   ($\%_{L}$) V. Low Factual &  %\cellcolor[rgb]{0.682,0.857,0.478} 
  0.43\ddag &                                 \cellcolor{gray!5} &                 \cellcolor{gray!5} &                              \cellcolor{gray!5} &                  \cellcolor{gray!5} &                                   \cellcolor{gray!5} &                  \cellcolor{gray!5} \\
 &     
 
  \textcolor{mean_frac_pnnl_deceptive}{\rule{0.2cm}{0.2cm}} % \cellcolor{mean_frac_pnnl_deceptive!15}
  ($\%_{L}$) Deceptive &  %\cellcolor[rgb]{0.700,0.868,0.499} 
  0.42\ddag &                                 \cellcolor{gray!5} &                 \cellcolor{gray!5} &                              \cellcolor{gray!5} &                  \cellcolor{gray!5} &                                   \cellcolor{gray!5} &                  \cellcolor{gray!5} \\
  
   \ignore{ % remove lowest & only #
 &      
  \textcolor{mean_num_pnnl_deceptive}{\rule{0.2cm}{0.2cm}} % \cellcolor{mean_num_pnnl_deceptive!15} 
  ($\#_{L}$) Deceptive  &   %\cellcolor[rgb]{0.782,0.908,0.622}
  0.33\dag &                                 \cellcolor{gray!5} &                 \cellcolor{gray!5} &                              \cellcolor{gray!5} &                  \cellcolor{gray!5} &                                   \cellcolor{gray!5} &                  \cellcolor{gray!5} \\
  }
  \midrule
\multicolumn{9}{l}{\footnotesize \it Recall}\\
  \midrule 
\multirow[c]{7}{*}{}&% R    &    
\cellcolor{gray!5} &                 \cellcolor{gray!5} & 

 \textcolor{frac_true_deceptive}{\rule{0.2cm}{0.2cm}}  %\cellcolor{frac_true_deceptive!15} 
 \cellcolor{white}
($\%_{T}$)  Deceptive &   %\cellcolor[rgb]{0.760,0.899,0.588} 
 0.35\dag &                              \cellcolor{gray!5} &                  \cellcolor{gray!5} &                                   \cellcolor{gray!5} &                  \cellcolor{gray!5} \\
\bottomrule
\end{tabular}
 
   \caption{
    %Correlation 
    Characteristics with at least moderate correlation (Pearson $|r|>0.3$) with model performance metrics across Communities or Authors. $\dag$ denotes a p-value $<.05$, $\ddag$ denotes a p-value $<.01$. ``$\%_T$'' refers to links in the test set and ``$\%_L$'' refers to links submitted to communities (Community characteristics) or by authors (Author characteristics) considering all posts submitted to \reddit. Colored squares correspond to color used in Figure~\ref{fig:number_line}.  
    }
    \label{tab:agg_correlations}
\end{table*}

\begin{figure*}
    \centering
    \vspace{-\baselineskip}
    \includegraphics[width=0.9\textwidth]{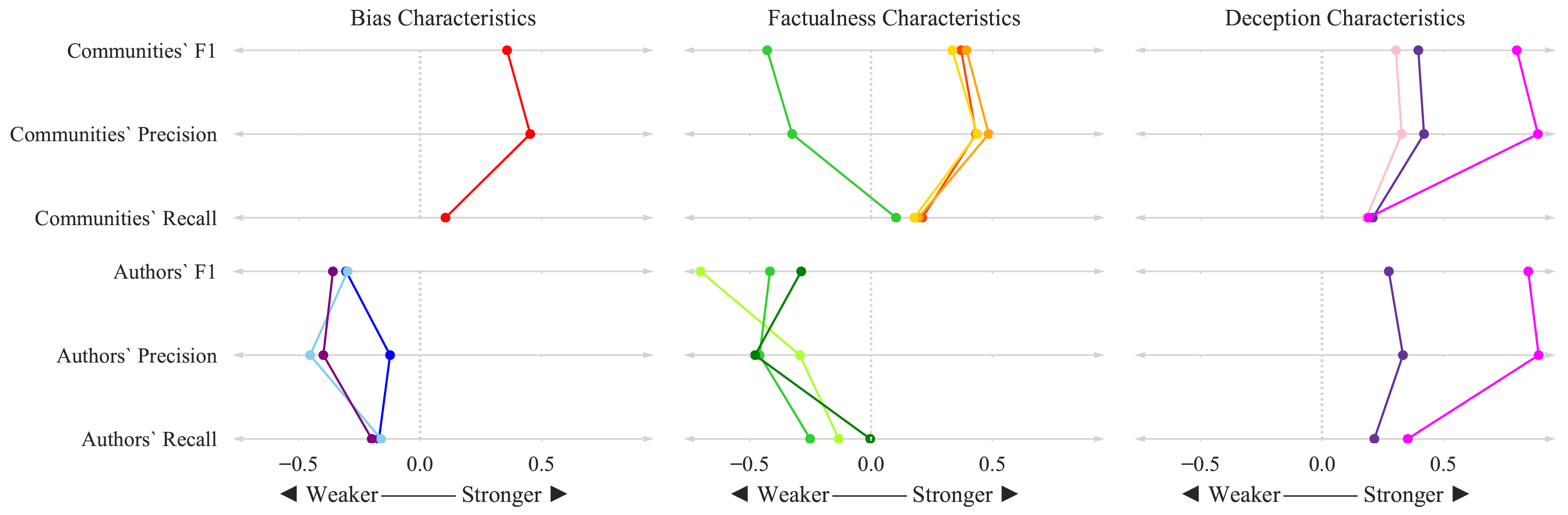} 
    %\small
\footnotesize 
%\hspace{6\baselineskip}
\begin{tikzpicture} 
\begin{axis}[%
hide axis, height =.75in,
xmin=0,xmax=50,ymin=0,ymax=0.4,
legend style={at={(0.5,1)},%0.5)},
	anchor=north,
	legend cell align=left,
	legend columns=1%legend columns=4,%2, %% Controls how many columns in the legend
	column sep=.2cm,draw=none},   
	every axis/.append style={
        legend image post style={xscale=.75}
    }
] 

\addlegendimage{no marks, mean_frac_left_bias, thick}
\addlegendentry{\% All Links = Left Bias}

\addlegendimage{no marks, mean_frac_center_left_bias, thick}
\addlegendentry{\% All Links = Center Left Bias 
}

\addlegendimage{no marks, mean_frac_center_bias, thick}
\addlegendentry{\% All Links = Center Bias}
    
\addlegendimage{no marks, mean_frac_extreme_right_bias, thick}
\addlegendentry{\% All Links = Extreme Right Bias}

\addlegendimage{empty legend}
\addlegendentry{}
\addlegendimage{empty legend}
\addlegendentry{}

\end{axis}
\end{tikzpicture} 
%\hspace{\baselineskip}
\begin{tikzpicture} 
\begin{axis}[%
hide axis, height =.75in,
xmin=0,xmax=50,ymin=0,ymax=0.4,
legend style={at={(0.5,1)},%0.5)},
	anchor=north,
	legend cell align=left,
	legend columns=1%legend columns=4,%2, %% Controls how many columns in the legend
	column sep=.2cm,draw=none},   
	every axis/.append style={
        legend image post style={xscale=.75}
    }
]

%\addlegendimage{empty legend}
%\addlegendentry[yshift=5pt]{\bf Factualness:}

\addlegendimage{no marks, mean_frac_very_low_factual, thick}
\addlegendentry{\% All Links = Very Low Factual}

\addlegendimage{no marks, mean_frac_low_factual, thick}
\addlegendentry{\% All Links = Low Factual}

\addlegendimage{no marks, mean_frac_mixed_factual, thick}
\addlegendentry{\% All Links = Mixed Factual}

%\addlegendimage{empty legend}
%\addlegendentry{}

\addlegendimage{no marks, mean_frac_mostly_factual, thick}
\addlegendentry{\% All Links = Mostly Factual}

\addlegendimage{no marks, mean_frac_high_factual, thick}
\addlegendentry{\% All Links = High Factual}

\addlegendimage{no marks, mean_frac_very_high_factual, thick}
\addlegendentry{\% All Links = Very High Factual}

\end{axis}
\end{tikzpicture} 
%\hspace{2\baselineskip}
\begin{tikzpicture} 
\begin{axis}[%
hide axis, height =.75in,
xmin=0,xmax=50,ymin=0,ymax=0.4,
legend style={at={(0.5,1)},%0.5)},
	anchor=north,
	legend cell align=left,
	legend columns=1%legend columns=4,%2, %% Controls how many columns in the legend
	column sep=.2cm,draw=none},   
	every axis/.append style={
        legend image post style={xscale=.75}
    }
]

%\addlegendimage{empty legend}
%\addlegendentry[yshift=5pt]{\bf Categorical:}

\addlegendimage{no marks, mean_num_pnnl_deceptive, thick}
\addlegendentry{\# All Links = Deceptive Links}

\addlegendimage{no marks, mean_frac_pnnl_deceptive, thick}
\addlegendentry{\% All Links = Deceptive}

\addlegendimage{no marks, frac_true_deceptive, thick}
\addlegendentry{\% Test = Deceptive}
%\addlegendentry{\% Test Links = True Deceptive}

\addlegendimage{empty legend}
\addlegendentry{}
\addlegendimage{empty legend}
\addlegendentry{}
 
\addlegendimage{empty legend}
\addlegendentry{}

\end{axis}
\end{tikzpicture}

\vspace{-\baselineskip}
    \caption{Correlation coefficients between characteristics and aggregated community/author performance metrics: F1, precision, and recall. All characteristics with an absolute Pearson's $r$ correlation coefficient greater than 0.3 for at least one metric are included. Generally, stronger model performance is correlated with more right bias, low factual, and deceptive content, while weaker performance is correlated with more left bias and high factual content. 
    }
    \label{fig:number_line}
\end{figure*}

To examine factors that explain the model's performance \textit{in aggregate}, we consider performance across individual authors and communities. First, { compute performance metrics (precision, recall, and F1 score) for the post} across posts by every author, and then correlate these metrics with authors' characteristics. We repeat this process for communities, as well.  
Characteristics with at least moderate correlation ($r\ge 0.3$) are presented in Table~\ref{tab:agg_correlations}.  
Compared to post-level correlations with model confidence, we immediately notice that both aggregated community- and author-level correlations are much stronger, \eg a maximum correlation value of 0.70 for features derived from all-\reddit data, compared to a maximum correlation value of 0.37 for individual posts. This observation suggests that model performance is more strongly correlated with characteristics across entire communities or authors rather than individual posts.

{{ For both authors and communities, the characteristics most strongly correlated with a higher F1 score is the fraction of deceptive content submitted in that community or by that author. These correlations are strongest (0.80 for communities, 0.85 for authors) when we examine just content from the test set, but are still substantial (0.42 and 0.33, respectively) when considering content across all of \reddit. }
Computing the fraction of deceptive posts in the test set for each community/author results in larger fractions than when considering all of \reddit, as the test set contains a greater proportion of deceptive posts than \reddit in general. 
We also note that while the characteristics most strongly correlated with F1-Score and Precision are quite similar to one another, there are almost no features which are at least moderately (\ie $>\pm .3$) correlated with recall. This aligns with our findings when comparing the \LSTMmodelname model to baseline performance (\sect\ref{fig:baseline_comparison}), where we found that author and community characteristics are more useful for achieving high precision than high recall.

Grouping the characteristics from Table~\ref{tab:agg_correlations} and displaying them visually, as in Figure~\ref{fig:number_line} allows us to easily distinguish the differences between ordinal characteristics such as bias (extreme left to extreme right) and factualness (very low to very high).  
Across both communities and authors, greater fractions of left bias posts are correlated with \textit{weaker} model performance, whereas greater fractions of right bias posts are correlated with \textit{stronger} model performance. Similarly, greater fractions of high factual posts are correlated with \textit{weaker} performance, while more low factual posts are correlated with \textit{stronger} model performance.

\subsection{Popularity and Community Acceptance} 
\label{sec:score_corrs}

Next, we consider whether our model performs equitably across posts that do and do not gain community acceptance, and across varying levels of popularity. 
We examine the correlation of each post's community-normalized score\footnote{Normalized by the median score for all posts from the same month in the same subreddit.} and the \LSTMmodelname model's confidence when predicting the true class annotation of the post. 
For the test set overall, this correlation is +0.094, but is higher for deceptive posts (+0.104) than for credible posts (+0.083). We found that all correlations are significant (p-values $< 10^{-5}$) but the effect is small.  

In Table~\ref{tab:score_decile_corrs}, we see that there are no significant correlations greater than .2 for posts with low to moderate community acceptance. { However, for the posts most highly accepted by the community (\ie those in the 9th and 10th deciles), the correlations are both significant and relatively strong.}  
{{ This suggests that in general, the model is more confident on posts that are more accepted by the community, but only for posts that are highly accepted by the community. We also compute the same correlation coefficients for posts linking to credible and deceptive news sources separately, and find the trend is magnified: For posts linking to deceptive sources that are most widely accepted within their given community, community acceptance is highly (+0.51 and +0.4) correlated with greater model confidence. In contrast, for posts linking to credible news sources that are strongly positively received or promoted by the community, the model is actually slightly \textit{less} confident (correlation coefficient of -.017). } This is an important distinction in behavior, particularly for deception detection models that may be leveraged as automated systems to flag deceptive content to investigate or intervene against or as a gate-keeping mechanism to slow the spread of misinformation online.

\begin{table}[t]
    \centering
    \small
    
\footnotesize
\begin{tabular}{p{.2cm}p{.2cm}c|rrr}
\toprule
{} & {} & {} &                                     \textbf{All} &                                 \textbf{Credible} &                               \textbf{Deceptive} \\
\midrule
\multirow{10}{*}{\rot{$\leftarrow$ More \hspace{20pt} Less $\rightarrow$}} & \multirow{10}{*}{\rot{\textbf{Community Acceptance}}}
        & \textbf{Decile 1 } &       \cellcolor[rgb]{0.975,0.946,0.961} $-0.05$ &       \cellcolor[rgb]{0.946,0.966,0.918} $+0.07$ &       \cellcolor[rgb]{0.965,0.968,0.960} $+0.01$ \\
{} & {} & \textbf{Decile 2 } &       \cellcolor[rgb]{0.959,0.968,0.948} $+0.02$ &       \cellcolor[rgb]{0.975,0.946,0.961} $-0.05$ &       \cellcolor[rgb]{0.957,0.967,0.942} $+0.03$ \\
{} & {} & \textbf{Decile 3 } &       \cellcolor[rgb]{0.971,0.960,0.966} $-0.02$ &       \cellcolor[rgb]{0.976,0.942,0.959} $-0.06$ &       \cellcolor[rgb]{0.954,0.967,0.936} $+0.04$ \\
{} & {} & \textbf{Decile 4 } &       \cellcolor[rgb]{0.974,0.949,0.962} $-0.04$ &       \cellcolor[rgb]{0.972,0.956,0.964} $-0.03$ &       \cellcolor[rgb]{0.973,0.953,0.963} $-0.03$ \\
{} & {} & \textbf{Decile 5 } &       \cellcolor[rgb]{0.971,0.960,0.966} $-0.02$ &       \cellcolor[rgb]{0.973,0.953,0.963} $-0.04$ &       \cellcolor[rgb]{0.967,0.968,0.966} $+0.01$ \\
{} & {} & \textbf{Decile 6 } &       \cellcolor[rgb]{0.949,0.966,0.924} $+0.06$ &       \cellcolor[rgb]{0.965,0.968,0.960} $+0.01$ &       \cellcolor[rgb]{0.946,0.966,0.918} $+0.07$ \\
{} & {} & \textbf{Decile 7 } &  \cellcolor[rgb]{0.933,0.964,0.888} $+0.10\ddag$ &       \cellcolor[rgb]{0.918,0.963,0.852} $+0.15$ &  \cellcolor[rgb]{0.923,0.963,0.864} $+0.14\ddag$ \\
{} & {} & \textbf{Decile 8 } &  \cellcolor[rgb]{0.981,0.921,0.952} $-0.11\ddag$ &       \cellcolor[rgb]{0.982,0.917,0.951} $-0.11$ &  \cellcolor[rgb]{0.905,0.961,0.822} $+0.19\ddag$ \\
{} & {} & \textbf{Decile 9 } &  \cellcolor[rgb]{0.638,0.828,0.425} $+0.47\ddag$ &  \cellcolor[rgb]{0.988,0.893,0.942} $-0.17\ddag$ &  \cellcolor[rgb]{0.594,0.800,0.372} $+0.51\ddag$ \\
{} & {} & \textbf{Decile 10} &  \cellcolor[rgb]{0.918,0.963,0.852} $+0.15\ddag$ &       \cellcolor[rgb]{0.962,0.968,0.954} $+0.02$ &  \cellcolor[rgb]{0.717,0.880,0.520} $+0.40\ddag$ \\
\bottomrule
\end{tabular}
    \caption{Correlations between community acceptance, by decile,  
    and the \LSTMmodelname model confidence (true class).  
    $\dag$ denotes a  
    p-value $< .05$, $\ddag$ for p-value $< .01$.
    }
    \label{tab:score_decile_corrs}
\end{table}

 %%%%%%%%%%%%%%%%%%%%

\section{Discussion and Conclusions}\label{sec:summary}
In summary, we quantify the context of deceptive and credible posts by computing community and author characteristics  
and 
use these characteristics, 
to explain and characterize the performance of an LSTM-based model for deception detection, examining {performance variance} across communities or users to identify characteristics of sub-populations where the model disproportionately underperforms.

We find that in general, sub-population characteristics are more strongly correlated with aggregate performance, and that, for both communities and authors, the model is more effective at identifying deceptive posts (higher F1 and precision) when the author/community has a greater tendency to submit or receive posts linked to deceptive, low factual, and right biased news sources. In contrast, a greater tendency to submit or receive posts linked to high factual and center biased content are correlated with weaker F1 and precision -- the model is more likely to fail when identifying posts submitted to communities or users that engage with more trustworthy news sources.

We also investigate the impact that community acceptance has on model performance, using community-normalized {scores} to quantify acceptance. We find that, for posts with low to moderator community acceptance, correlations with the model's confidence that a post belongs to its groundtruth annotation class are small, but for posts that are strongly accepted by the community they are submitted to, acceptance is strongly correlated with increased model confidence for deceptive content, but only moderately correlated with \textit{decreased} model confidence for credible content. It is important to consider what kinds of failures are most impactful given the specific application of a model. For example, if considering a deception detection model for use as an intervention strategy, it may be more important for a model to have greater reliability when identifying content that gains widespread community acceptance or popularity as we find our \LSTMmodelname model does --- this is an important direction of evaluation for researchers in the deception detection domain to consider.

We encourage NLP researchers working in the deception detection space to look beyond overall test-set performance metrics such as F1 score. Although many models achieve high overall F1 score, the performance of these models varies dramatically from community to community. 
Decisions about model design and training should not be made without considering the intended application of the model. For example, a model tasked with flagging posts for human review may be optimized with a very different precision-recall trade-off than a model tasked with automatically taking entire enforcement actions, such as removing content.

 %%%%%%%%%%%%%%%%%%%%

\section*{Acknowledgements}
 This research was supported by the Laboratory Directed Research and Development Program at Pacific Northwest National Laboratory, a multi-program national laboratory operated by Battelle for the U.S. Department of Energy.

\bibliographystyle{acl_natbib}
%\bibliography{bibliography}

\end{document}